\documentclass[12pt]{article}
\usepackage{amssymb,amsmath,epsfig}

\begin{document}

\title{\bf Gravitational Collapse: Expanding and Collapsing Regions
\thanks{Talk at {\it The Second Italian-Pakistani Workshop on Relativistic Astrophysics},
held at ICRANet Center in Pescara, July 8-10, 2009.}}
\author{M. Sharif \thanks{msharif@math.pu.edu.pk} and G. Abbas
\thanks{abbasg91@yahoo.com}\\
Department of Mathematics, University of the Punjab,\\
Quaid-e-Azam Campus, Lahore-54590, Pakistan.}

\date{}
\maketitle

\begin{abstract}
We investigate the expanding and collapsing regions by taking two
well-known spherically symmetric spacetimes. For this purpose, the
general formalism is developed by using Israel junction conditions
for arbitrary spacetimes. This has been used to obtain the surface
energy density and the tangential pressure. The minimal pressure
provides the gateway to explore the expanding and collapsing
regions. We take Minkowski and Kantowski-Sachs spacetimes and use
the general formulation to investigate the expanding and
collapsing regions of the shell.
\end{abstract}


\section{Introduction}

The final outcome of the gravitational collapse in General
Relativity (GR) is an issue of great importance from the
perspective of black hole physics and its astrophysical
implications. A continual gravitational collapse without any final
equilibrium state, would end as a spacetime singularity. The
singularity theorems indicate the existence of singularities in
the form of either future or past incomplete timelike geodesics.

The main motivation for researchers working in the field of
gravitational collapse is to determine the form of spacetime
singularity (clothed or naked), resulting from the collapse of
massive objects. For this purpose, Penrose \cite{1} suggested a
hypothesis known as \textit{Cosmic Censorship Hypothesis} (CCH)
which states that in generic situation all singularities arising
from regular initial data are clothed by event horizon and hence
invisible to distant observers. Over the past three decades, this
conjecture remains un-proved and un-solved problem in the theory and
applications of black hole physics.

After the failure of many attempts to establish CCH, Penrose
\cite{2} concluded that continual collapse of a massive object would
end as black hole or naked singularity, depending on the initial
conditions and equation of state. Virbhadra et al. \cite{3}
introduced a new theoretical tool using the gravitational lensing
phenomena to discuss the spacetime singularity. Also, Virbhadra and
Ellis \cite{4} classified the naked singularity by the gravitational
lensing into two kinds: weak naked singularity (those contained
within at least one photon sphere \cite{5}) and strong naked
singularity (those not contained within any photon sphere).
Virbhadra \cite{6} explored the useful results to investigate the
Seifert's conjecture for the naked singularity. He also found that
naked singularity forming in the Vaidya null dust collapse supports
the Seifert's conjecture \cite{7}. The same author \cite{8} used the
gravitational lensing phenomena to find an improved form of the CCH.

Oppenheimer and Snyder \cite{9} are the pioneers who studied the
dust collapse by taking the static Schwarzschild spacetime as an
exterior and Friedmann like solution as an interior spacetime.
They found black hole as end state of the gravitational collapse.
This work was generalized by Markovic and Shapiro \cite{10} for
positive cosmological constant.

To study the gravitational collapse, it is of vital importance to
consider an appropriate geometry of the interior and exterior
regions and junction conditions which allows the smooth matching of
these regions \cite{11}. Qadir with his collaborators \cite{12}
investigated the use of junction conditions for collapse of black
hole in a closed Friedmann universe which was later extended with
positive and negative cosmological constants. Lake \cite{15}
extended the work of Oppenheimer and Snyder for positive and
negative cosmological constants. Debnath et al. \cite{16} explored
the quasi-spherical collapse with cosmological constant by using the
Israel junction conditions modified by Santos. Ghosh and Deshkar
\cite{17} discussed the higher dimensional spherically symmetric
dust collapse. Sharif and Ahmad \cite{18} extended the spherically
symmetric gravitational collapse with positive cosmological constant
for perfect fluid. Nath et al. \cite{19} studied the gravitational
collapse of non-viscous, heat conducting fluid in the presence of
electromagnetic field. In a recent paper \cite{20}, we have
investigated the effect of electromagnetic field on the perfect
fluid collapse by using the junction conditions.

Villas da Rocha et al. \cite{21} discussed the self-similar
gravitational collapse of perfect fluid using Israel's method.
Pereira and Wang \cite{22} studied the gravitational collapse of
cylindrical shells made of counter rotating dust particles by
using the same analysis. Sharif and Ahmad \cite{23} extended
this work to plane symmetric spacetime. Recently, Sharif and Iqbal
\cite{24} have investigated the spherically symmetric
gravitational collapse for a class of metrics. In this paper, we
generalize this work. We develop a general formulation and apply it
to two well-known spherically symmetric spacetimes (Minkoski and
Kantowski-Sachs) to study the gravitational collapse.

The main objective of this work is to investigate the expanding and
collapsing regions in the framework of these well-known spacetimes.
The plan of the paper is as follows: In the next section, the
general formalism is presented. In section \textbf{3}, we discuss
the application of this formalism to the known spacetimes. We
conclude our discussion in the last section. All the Latin and Greek
indices vary from 0 to 3, otherwise it will be mentioned.

\section{General Formulation for Surface Energy-Momentum Tensor}

A spacelike $3D$ hypersurface $\Sigma$ is taken which separates
 the interior  and exterior regions of a star represented by $4D$ manifolds
 $M^-$ and $M^+$ respectively. For interior region $M^-$, we
take spherically symmetric spacetime given by
\begin{equation}\label{1}
ds_-^2=W^-dt^2-X^-dr^2-Y^-(d\theta^2+\sin\theta^2d\phi^2),
\end{equation}
where $W^-,~X^-$ and $Y^-$ are functions of $t$ and $r$. For the
exterior manifold $M^+$, we take the line element of the form
\begin{equation}\label{2}
ds_+^2=W^+dT^2-X^+dR^2-Y^+(d\theta^2+\sin\theta^2d\phi^2),
\end{equation}
where $W^+,~X^+$ and $Y^+$ are functions of $T$ and $R$.

It is assumed that interior and exterior spacetimes are smoothly
matched on the hypersurface $\Sigma$ by the continuity of line
elements over $\Sigma$ following Israel junction conditions
\cite{11}. Thus we have
\begin{equation}\label{3}
(ds^2_-)_{\Sigma}=(ds^2_+)_{\Sigma}=ds^2_{\Sigma}.
\end{equation}
The equations of hypersurface in terms of interior and exterior
coordinates are
\begin{eqnarray}\label{4}
f_-(r,t)=r-r_{0}(t)=0,\\ \label{5}
{f_+(R,T)=R-R_{0}(T)=0}.
\end{eqnarray}
When we make use of these equations, we obtain the following
interior and exterior metrics
\begin{eqnarray}\label{6}
(ds_-^2)_\Sigma
&=&[W^-(t,r_0(t))-X^-(t,r_0(t)){r'_0}^2(t)]dt^2\\
&-&Y^-(t,r_0(t))(d\theta^2+\sin\theta^2d\phi^2),\nonumber\\
\label{7}
(ds_+^2)_\Sigma&=&[W^+(T,R_0(T))-X^+(T,R_0(T)){R'_0}^2(t)]dT^2\nonumber\\
&-&Y^+(T,R_0(T)) (d\theta^2+\sin\theta^2d\phi^2),
\end{eqnarray}
where prime means derivative with respect to the indicated variable.

We define the intrinsic metric on the hypersurface as follows
\begin{equation}\label{8}
(ds^2)_\Sigma=d\tau^2-Y(\tau)(d\theta^2+\sin\theta^2d\phi^2),
\end{equation}
where $\tau$ is the proper time. Using Eqs.(\ref{6})-(\ref{8}) in
(\ref{3}), we get
\begin{eqnarray}\label{9}
d\tau&=&[W^-(t,r_0(t))-X^-(t,r_0(t)){r'_0}^2(t)]^\frac{1}{2}dt\nonumber\\
&=&[W^+(T,R_0(T))-X^+(T,R_0(T)){R'_0}^2(t)]^\frac{1}{2}dT,\\
\label{10a} Y(\tau)&=&Y^-(t,r_0(t))=Y^+(T,R_0(T)).
\end{eqnarray}
Also, from Eqs.(\ref{4}) and (\ref{5}), the outward unit normals
(timelike) in the coordinates of $M^-$ and $M^+$, are given by
\begin{eqnarray}\label{11}
n^-_\mu&=&[\frac{W^- X^-}{W^- -X^-{r'_0}^2(t)}]^\frac{1}{2}(-r'_0(t),1,0,0),\\
\label{12} n^+_\mu&=&[\frac{W^+ X^+}{W^+
-X^+{R'_0}^2(T)}]^\frac{1}{2}(-R'_0(T),1,0,0).
\end{eqnarray}

Now the extrinsic curvature $K_{ij}$ is defined as
\begin{equation}\label{13}
K^{\pm}_{ij}=n^{\pm}_{\sigma}(\frac{{\partial}^2x^{\sigma}_{\pm}}
{{\partial}{\xi}^i{\partial}{\xi}^j}+{\Gamma}^{\sigma}_{{\mu}{\nu}}
\frac{{{\partial}x^{\mu}_{\pm}}{{\partial}x^{\nu}_{\pm}}}
{{\partial}{\xi}^i{\partial}{\xi}^j}),\quad({i},{j}=0,2,3).
\end{equation}
Here $\xi^i$ correspond to the coordinates on ${\Sigma }$,
$x^{\sigma}_{\pm}$ stand for coordinates in $M^{\pm}$,  the
Christoffel symbols $\Gamma^{\sigma}_{{\mu}{\nu}}$ are calculated
from the interior or exterior spacetimes and $n^{\pm}_{\sigma}$ are
components of outward unit normals to ${\Sigma}$ in the coordinates
$x^{\sigma}_{\pm}$. The non-vanishing components of ${K^+}_{ij}$ (exterior region)
are given by
\begin{eqnarray}\label{14}
K^+_{\tau\tau}&=&\frac{(W^+X^+)^\frac{1}{2}}
{[{W^{+}-X^+{R'_0}^2(T)}]^\frac{3}{2}}[{R''_0}(T)+\frac{{W^+}_{,R}}{W^+}
+\frac{{X^+}_{,T}}{X^+}R'_0(T)\nonumber\\
&+&\frac{{X^+}_{,R}}{2X^+}{R'}^2_0(T)-\frac{{W^+}_{,T}}{2W^+}{R'}_0(T)
-\frac{{X^+}_{,T}}{2W^+}{R'}^3_0(T)-\frac{{W^+}_{,T}}{W^+}{R'}^2_0(T)], \\
\label{15}
K^+_{\theta\theta}&=&\frac{1}{2}[\frac{(W^+X^+)}{W^{+}-X^+{R'_0}^2(T)}]^\frac{1}{2}
[-\frac{{Y^+}_{,R}}{X^+}-\frac{{Y^+}_{,T}}{W^+}R'_0(T)],\\
\label{16} K^+_{\phi\phi}&=&\sin^2\theta K^+_{\theta\theta}.
\end{eqnarray}
The non-vanishing components of extrinsic curvature, ${K^-}_{ij}$,
in terms of interior coordinates are found from the above expression
by replacing
\begin{equation}\label{17}
W^+,~X^+,~Y^+,~R_0(T),~T,~R~\rightarrow~W^-~X^-,~Y^-,~r_0(t),~t,~r.
\end{equation}

The surface energy-momentum tensor in terms of ${K}^\pm_{ij}$ and
$\gamma_{ij}$ can be defined as \cite{11}
\begin{equation}\label{18}
S_{ij}=\frac{1}{\kappa}\{[{K}_{ij} ]-\gamma_{ij}[K]\},
\end{equation}
where ${\kappa}$ is coupling constant and $\gamma_{ij}$ represents
the metric coefficients of the hypersurface $\Sigma$. Also, we have
\begin{equation}\label{19}
[K_{ij}]={{K}^+}_{ij}-{{K}^-}_{ij}, \quad[K]=\gamma^{ij}[K_{ij}].
\end{equation}
Using Eqs.(\ref{14})-(\ref{16}) and the corresponding expression
for ${{K}^-}_{ij}$, we can express $S_{ij}$ in the form
\begin{equation}\label{20}
S_{ij}=\rho{\omega}_i{\omega}_j+ p({\theta}_i{\theta}_j+\phi_i
\phi_j),  \quad (i,j=\tau,\theta,\phi),
\end{equation}
where $\rho$ is the surface energy density, $p$ is the tangential
pressure provided that they satisfy some energy conditions
\cite{25} and $\omega_i,~\theta_i,~\phi_i$ are unit vectors
defined on the surface given by
\begin{equation}\label{21}
\omega_i ={\delta^\tau}_i,\quad
\theta_i=Y^\frac{1}{2}{\delta^\theta}_i,\quad
\phi_i=Y^\frac{1}{2}\sin^2\theta{\delta^\phi}_i.
\end{equation}
The energy density $\rho$ and pressure $p$ are given by
\begin{equation}\label{22}
\rho =\frac{2}{{\kappa} Y}[K_{{\theta\theta}}],\quad p
=\frac{1}{\kappa}
\{[K_{\tau\tau}]-\frac{[{K_{\theta\theta}}]}{Y}\}.
\end{equation}

\section{Application}

Now we use this formulation by taking particular interior and
exterior spacetimes. We take Minkowski spacetime as an interior and
Kantowski-Sach spacetime as an exterior region. The Minkowski
spacetime is
\begin{equation}\label{23}
ds_-^2=dt^2-dr^2-r^2(d\theta^2+\sin\theta^2d\phi^2).
\end{equation}
The Kantowski-Sach spacetime is given by
\begin{equation}\label{24}
ds_+^2=dT^2-{A^2(T)}dR^2-{B^2(T)}(d\theta^2+\sin\theta^2d\phi^2).
\end{equation}
The junction conditions (\ref{9}) and (\ref{10a}) can be found from
the above spacetimes as
\begin{eqnarray}\label{25}
d\tau&=&[1-{r'_0}^2(t)]^\frac{1}{2}dt
=[1-A^2(T){R'_0}^2(T)]^\frac{1}{2}dT,\\\label{26}
r_0(t)&=&B(T).
\end{eqnarray}
These equations yield
\begin{equation}\label{27}
(\frac{dT}{dt})^2=\frac{1}{[1-A^2(T){R'_0}^2(T)+{B'}^2(T)]}\equiv
\frac{1}{{\Delta}^2}.
\end{equation}

From Eqs.(\ref{26}) and (\ref{27}), one gets
\begin{eqnarray}\label{29}
r''_0(t)&=&\frac
{B''(T)}{\Delta^2}+\frac{1}{\Delta^4}[B'(T)(A^2(T){R'_0}(T){R''_0}(T)\nonumber\\
&+&A(T){R'_0}^2(T)A'(T)-B'(T)B''(T))].
\end{eqnarray}
Using Eqs.(\ref{26})-(\ref{29}) in the extrinsic curvature
components, we get the expression for $\rho$ and $p$ in the
following form
\begin{eqnarray}
\label{30}
\rho&=&\frac{2}{\kappa}\frac{[\Delta-A(T)B'(T){R'_0}(T)]}
{(1-A^2(T){R'_0}^2(T))^\frac{1}{2}},\\
\label{31}p&=&\frac{1}{{\kappa}\Delta{[1-(A(T)R(T))^2]}^\frac{3}{2}}
[\Delta(A(T){R''_0}(T)+2{R'_0}
(T)A'(T)\nonumber\\
&-&A^2(T){R'_0}^2(T)A'(T))-B''(T)(1-A^2(T){R'_0}^2(T))\nonumber\\
&-&B'(T)A(T){R'_0}(T)(A(T){R''_0}(T)+A'(T){R'_0}(T))\nonumber\\
&-& \Delta(1-A^2(T){R'_0}^2(T))(\Delta-A(T)B'(T){R'_0}(T))].
\end{eqnarray}
Now for the minimum effects of shell on the collapse, substitute
$p=0$ in the above equation which gives
\begin{eqnarray}\label{32}
R''_0(T)&=&\frac{{R'_0}(T)}{\Delta A(T)-B'(T)R'(T)A^2(T)}
[(B''(T)-B'(T)-1)\nonumber\\
&\times&(1-A^2(T){R'_0}^2(T))+\Delta(A(T){B'}(T)\nonumber\\
&+&A^2(T)B^2(T){R'_0}^2(T)+2A(T))-B'(T)A^2(T)].
\end{eqnarray}

For the analysis of gravitational collapse, it is necessary to solve
this equation for ${R'_0}(T)$. The exact solution of this equation
is too complicated to provide any insight. However, it would be
interesting to consider a particular case for which this equations
reduces to the following form
\begin{equation}\label{33}
R''_0(T)={R'_0}(T)\alpha,
\end{equation}
where we have assumed $\alpha$ to be an arbitrary constant given by
\begin{eqnarray}\label{34}
\alpha&=&\frac{1}{{\Delta
A(T)-B'(T)R'(T)A^2(T)}}[(B''(T)-B'(T)-1)(1-A^2(T){R'_0}^2(T))\nonumber\\
&+& \Delta(A(T){B'}(T)
+A^2(T)B^2(T){R'_0}^2(T)+2A(T))-B'(T)A^2(T)].
\end{eqnarray}
Integrating Eq.(\ref{33}), it follows that
\begin{equation}\label{35}
R'_0(T)= e^{{\alpha T+c_1}},
\end{equation}
where $c_1$ is constant of integration which may be positive or
negative. Thus we can take
\begin{equation*}
(1)\quad \alpha>0 \quad (2)\quad\alpha<0
\end{equation*}

\subsection {Case 1}

Here, we take $\alpha=1$ and $c_1=-1$, then Eq.(\ref{35}) takes the
form
\begin{equation}\label{36}
R'_0(T)= e^{{ T-1}}.
\end{equation}
Integration of this equation yields
\begin{equation}\label{37}
R_0(T)= e^{{ T-1}}+ c_2,
\end{equation}
where $c_2$ is another constant of integration. When $c_2=-1$, we
obtain from Eqs.(35) and (36) as follows
\begin{eqnarray}
R_0(T)&=&\left \{ \begin{array}{lll}
0&,&T=1,\nonumber\\
+\infty&,&T=+\infty,
\end{array}\right.\nonumber\\
R'_0(T)&=&\left \{ \begin{array}{lll}
1&,&T=1,\\
+\infty&,&T=+\infty
\end{array}\right.
\end{eqnarray}
We see from figures 1 and 2 that expansion starts at $T=1$ (radial
velocity is 1 and radial displacement is 0) and ends at $T=\infty$
(both radial velocity and displacement becomes infinite). In this
case, acceleration is positive which goes on increasing with the
passage of time and future directed spacelike geodesics exist in
this region. It is mentioned here that we have determined the
physical time interval by taking the integration constant as minus
times $\alpha$. All the positive values of $\alpha$ and negative
values of integration constants will lead to the same range of
radial velocity, displacement and time. This case gives the
expanding process.
\begin{figure} \center\epsfig{file=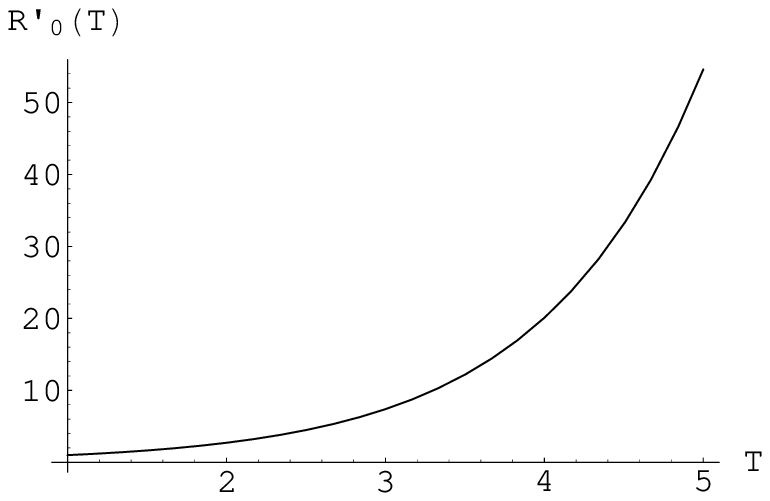,
width=0.4\linewidth}\caption{velocity-time graph}
\center\epsfig{file=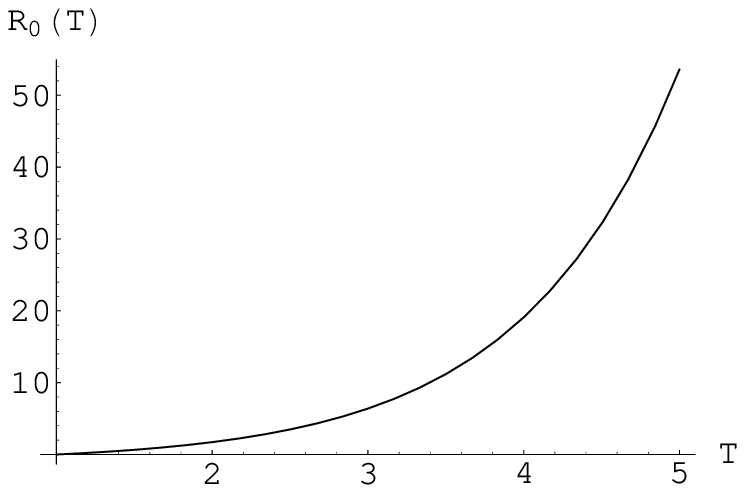,
width=0.4\linewidth}\caption{displacement-time graph}
\end{figure}

\subsection {Case 2}

\begin{figure} \center\epsfig{file=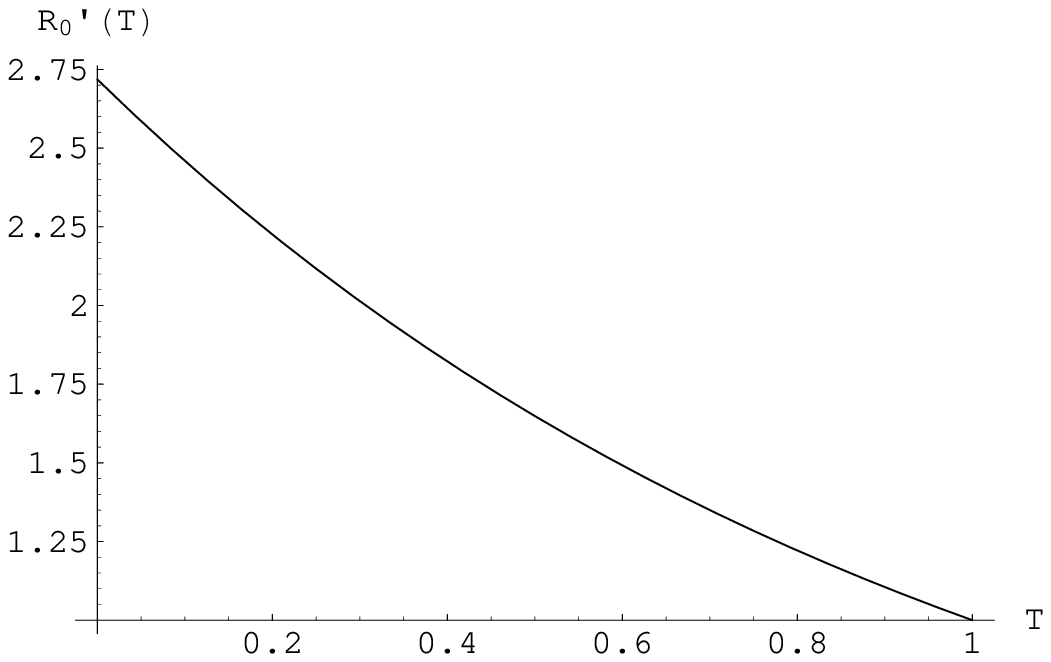,
width=0.4\linewidth}\caption{velocity-time graph}
\center\epsfig{file=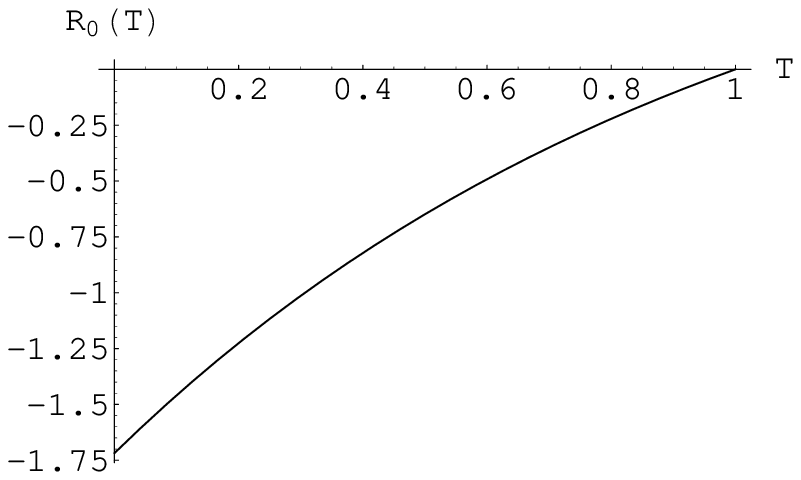,
width=0.4\linewidth}\caption{displacement-time graph}
\end{figure}
In this case, we take $\alpha=-1$ and $c_1=1$, then Eq.(\ref{35})
becomes
\begin{equation}\label{39}
R'_0(T)= e^{{ -T+1}}.
\end{equation}
Integrating this equation, it follows that
\begin{equation}\label{40}
R_0(T)=-e^{{-T+1}}+ c_3,
\end{equation}
where $c_3$ is another constant of integration. For $c_3=1$, we
obtain from Eqs.(\ref{39}) and (\ref{40})
\begin{eqnarray}
R_0(T)&=&\left \{\begin{array}{lll}
-1.7182&,&T=0,\nonumber\\
0&,&T=1,
\end{array}\right.\nonumber\\
R'_0(T)&=&\left \{ \begin{array}{lll}
+2.7182&, &T=0,\\
1&,       &T=1.
\end{array}\right.
\end{eqnarray}

From figures 3 and 4, it is obvious that shell starts collapsing at
$T= 0$ with radial velocity $2.7182$ and magnitude of displacement
$1.7182$ and ends with radial velocity $1$ and displacement $0$ at
$T=1$. Thus the time interval for the collapsing region is $0\leq
T\leq 1$. The acceleration decreases positively in this time
interval and past directed spacelike geodesics exist in this
region. The relation between $\alpha$ and integration constants
enable us to determine the physical time interval. All the negative
values of $\alpha$ and positive values of integration constants lead
to the same range of radial velocity, displacement and time. This
case represents the collapsing process.

\section{Summary and Conclusion}

In this paper, we investigate the expanding and collapsing regions
in the spherically symmetric background. For this purpose, we have
formulated a general formalism of surface energy-momentum tensor
using Israel junction conditions with arbitrary spacetimes. Further,
we have taken two particular spacetimes representing the interior
and exterior regions of a star. The surface density and pressure are
calculated by applying the general formulation with particular
spacetimes. To examine the minimal effects of shell on the collapse,
we assume that the tangential pressure $p=0$, which provides the
gateway for studying the expanding and collapsing regions.

After applying the assumption of minimal effect of shell on the
collapse, we obtain a simplest form given in Eq.(32). The arbitrary
constant $\alpha$ in Eq.(32) is either positive or negative. If we
take the constants of integration $c_1,c_2<0$ in Eqs.(34) and (36)
respectively then for all $\alpha>0$, we obtain the same time
interval, radial velocity and displacement for the expanding
regions. It is found that $\forall ~\alpha>0$ and $c_1=c_2=-\alpha$,
the expansion starts at $T=1$ and ends at $T=+\infty$. In the
expanding region, radial velocity ranges from $1$ to $+\infty$ and
displacement varies from $0$ to $+\infty$ while future directed
spacelike geodesics exist in this region .

On the other hand, all $\alpha<0$ and the constants of integration
$c_1,c_2>0$ in Eqs.(34) and (39) give the same time interval, radial
velocity and displacement in the collapsing regions. It is found
that $\forall~\alpha<0$ and $c_1=c_3=-\alpha$, the collapsing starts
at $T=0$ and ends at $T=1$. In the collapsing region, radial
velocity decreases from $2.7182$ to $1$ and magnitude of
displacement decreases $1.7182$ to $0$. Further, past directed
spacelike geodesics exist in this region. We find that in the
region, where collapse occurs, density remains finite which means
that collapse does not end as singularity and hence known as
non-singular collapse. One can extend this analysis by excluding the
assumption of minimal effects of shell on the collapse (i.e.,
$p\neq0$).

\vspace{0.25cm}

{\bf Acknowledgment}

\vspace{0.25cm}

One of us (MS) would like to thank the organizers for providing full
support to participate in \emph{II Italian-Pakistani Workshop on
Relativistic Astrophysics}. GA would like to thank the Higher
Education Commission, Islamabad, Pakistan for its financial support
through the {\it Indigenous Ph.D. 5000 Fellowship Program Batch-IV}.


\begin{thebibliography}{40}

\bibitem{1} Penrose, R.: Riv. Nuovo Cimento \textbf{1}(1969)252.

\bibitem{2} Penrose, R.: Phys. Rev. Lett.\textbf{ 14}(1964)57.

\bibitem{3} Virbhadra, K.S., Narasimha, D. and Chitre, S.M.: Astron. Astrophys.
\textbf{337}(1998)1.

\bibitem{4} Virbhadra, K.S., and Ellis, G.F.R.: Phys. Rev.
\textbf{D65}(2002)103004.

\bibitem{5} Paschoff, J.M.: \emph{Contemporary Astronomy}
(Harcourt Brace College Publishers, 1981).

\bibitem{6} Virbhadra, K.S.: Phys. Rev.
\textbf{D60}(1999)104041.

\bibitem{7} Seifert, H.: Proc. American Math. Soc. \textbf{1}(1950)287.

\bibitem{8} Virbhadra, K.S.: Phys. Rev.
\textbf{D79}(2009)083004.

\bibitem{9} Oppenhiemer, J.R. and Snyder, H.: Phys. Rev. \textbf{56}(1939)455.

\bibitem{10} Markovic, D. and Shapiro, S.L.: Phys. Rev.
\textbf{D61}(2000)084029.

\bibitem{11} Israel, W.: Nuovo Cimento \textbf{B44}(1966)1;
\emph{ibid.} \textbf{B48}(1967)463(\textbf{E}).

\bibitem{12} Qadir, A. and Wheeler, J.A.: \emph{ From SU(3) to Gravity: Yuval Ne'eman
Festschrift}, eds. Gotsman, E.S. and Tauber, G. (Cambridge
University Press, 1985); Qadir, A.: \emph{Proc. Fifth Marcel
Grossman Meeting}, eds. Blair, D.G. and Buckingham, M.J. (World
Scientific, 1989)593; Khan, I. and Qadir, A.: Lett. al. Nuovo
Cimento \textbf{41}(1984)493.

\bibitem{15} Lake, K.: Phys. Rev. \textbf{D62}(2000)027301.

\bibitem{16} Debnath, U., Nath, S. and Chakraborty, S.: Mon. Not. Roy. Astron. Soc.
\textbf{369}(2006)1916.

\bibitem{17} Ghosh, S.G. and Deshkar, D.W.: Astrophys. Space Sci.
\textbf{310}(2007)111.

\bibitem{18} Sharif, M. and  Ahmad, Z.: Mod. Phys. Lett.
\textbf{A22}(2007)1493., \emph{ibid.} 2947; J. Korean Phys. Society
\textbf{52}(2008)980; Acta Phys. Polonica B \textbf{39}(2008)1337.

\bibitem{19} Nath, S.,  Debnath, U. and Chakraborty, S.: Astrophys. Space Sci.
\textbf{313}(2008)431.

\bibitem{20}Sharif, M. and Abbas, G.: Mod. Phys. Lett.
\textbf{A24}(2009)2551.

\bibitem{21} Villas da Rocha, J.F., Wang, A. and
Santos, N.O.: Phys. Lett. {\bf A255}(1999)213.

\bibitem{22} Pereira, P.R.C.T. and Wang, A.: Phys. Rev. {\bf
D62}(2000)124001; Erratum {\it ibid.} {\bf D67}(2003)129902.

\bibitem{23} Sharif, M. and  Ahmad, Z.: Int. J. Mod. Phys.
\textbf{A23}(2008)181.

\bibitem{24} Sharif, M. and Iqbal, K.: Mod. Phys. Lett.
\textbf{A24}(2009)1533.

\bibitem{25} Hawking, S.W. and Ellis, G.F.R.: \emph{The Large Scale Structure of Spacetime}
(Cambridge University Press, Cambridge 1975).

\bibitem{26} MacCallaum,M.A.H.: \emph{In General Relativity: An Einstein Centtenary
Survey}, eds. Hawking, S. and Israel, W. (Cambridge University
Press, Cambridge, 1979).

\end{thebibliography}
\end{document}